\documentstyle[12pt,epsf]{article}

\def\ba{\begin{eqnarray}\samepage}
\def\ea{\end{eqnarray}}
\input amssym.def
\input amssym.tex

\newcommand{\news}{\setcounter{equation}{0}}
\newcommand{\diff}{\partial}
\newcommand{\bdiff}{{\bar{\partial}}}

\newcommand{\be}{\begin{equation}}
\newcommand{\ee}{\end{equation}}
\newcommand{\ben}{\begin{eqnarray}\displaystyle}
\newcommand{\een}{\end{eqnarray}}

\begin{document}
\title{Exact Non-Abelian Duality}

\author{Stephen Hewson\thanks{email sfh10@damtp.cam.ac.uk} \\
\&\\Malcolm Perry\thanks{email malcolm@damtp.cam.ac.uk} \\ \\D.A.M.T.P.\\University of
Cambridge\\Silver Street\\Cambridge CB3 9EW\\U.K.}

\maketitle

\begin{abstract}

We investigate non-abelian gaugings of  WZNW
models. When the gauged group is semisimple we are able to present
exact formulae for the dual conformal field theory, for all values of the
level $k$. The results are then applied to non-abelian target space duality in string theory,
showing that the standard formulae are quantum mechanically well
defined in the low energy limit if the gauged group is semisimple.

\end{abstract}

\newpage
\eject
\renewcommand{\thepage}{ }
\pagebreak

\renewcommand{\thepage}{\arabic{page}}

\pagestyle{plain}
\section{Introduction}
\news

One of the properties displayed by string theories which does not have
a point particle analogue is that of target space duality, or
T-duality. T-duality provides a method, valid to all orders in
perturbation theory, of relating seemingly inequivalent string
vacua. In the standard form, T-duality may be used to give a dual
description to any string theory where the background target space has a
group of isometries. In the case where the group of isometries is
abelian the theory is well understood \cite{gpr} and it is accepted that the
T-duality is a true symmetry of the string theory: when quantum effects are
taken into consideration, solutions which are dual to each other are
to be identified in the string moduli space. It is possible to
generalise the procedure to the case where the symmetry group is
$\it{non}$-${\it abelian}$\cite{qo,t,ks}. There is a problem  with this
generalisation however: although it is  possible to
generate new dual solutions of the low energy classical string
theory, the question
remains as to whether the solutions generated are truly equivalent
from a quantum mechanical point of view.

All the work to date in this area showed that if one took a consistent set of low
energy background fields, then  the duality transformation yields, in
certain situations, another set of low
energy background fields. Specifically, when the gauged subgroup is
semisimple, the new modes satisfy the first order beta-function
equations \cite{t}.

All this shows is that a low energy background will produce another
low energy background. If we wish to consider non-abelian duality to
be an exact quantum mechanical symmetry of the perturbative theory, as in the case of
abelian T-duality \cite{gpr}, then we require that it form a mapping from an
{\it{exact}} conformal string background to another exact background.

If it were the case that the dual models were quantum mechanically
equivalent then we would need to identify the original and dual
solutions in the string moduli space. This would give  non-abelian
duality the same standing as abelian duality, and would, of course,
further the search for equivalent string vacua.

We now address this problem and seek to implement the duality procedure
exactly 
at the level of the underlying conformal field theory. After dualising
we can take the low energy limit, to give us the correct low energy
theory.

If the non-abelian duality procedure is well defined quantum
mechanically then the order of the limits should not matter. If the procedure is not well defined then we will obtain anomaly terms
which are not present for the classical duality procedure.

\ben \mbox{CFT}\hskip0.5cm&\stackrel{limit}\longrightarrow&\hskip0.3cm\mbox{Low energy
effective action}\nonumber \\
\downarrow\hskip0.8cm&  \ \ &\hskip2.5cm\downarrow\nonumber \\
\mbox{Exact dual}& \ \ &\hskip1.5cm\mbox{Anomalies?} \nonumber \\
\downarrow\hskip0.8cm& \ \ &\hskip2.5cm\downarrow \nonumber \\
\mbox{Dual CFT}&\stackrel{limit}\longrightarrow&\mbox{Dual low energy
effective action}\,
\een

Since target space duality is, by definition, implemented at the level
of the $\sigma$-model we consider the exact result obtained under the
duality procedure by investigating WZNW
models. These models are exact conformal field theories and the
target spaces they may be thought of as representing are exact string
backgrounds. We show that we can find an explicit form of the  exact
dual of an exact string solution in the case when  the symmetry
group is semisimple. The formulae obtained reduce to the original
target space duality
equations in the low energy field theoretic limit,  which implies that
the original duality equations \cite{qo,t} are well defined quantum mechanically
if we consider gauging a semisimple isometry.

\section{String $\sigma$-models}
\news

We define the string theory via the conformally invariant 
 non-linear $\sigma$-model:

\be \label{nlsm}
S_{nlsm}={1\over{4\pi\alpha'}}\int
d^2\xi\left({\sqrt \gamma}\gamma^{ab}g_{MN}(X)\partial_a
X^M\partial_bX^N +\epsilon^{ab}B_{MN}(X)\partial_a
X^M\partial_bX^N \right) \,
\ee
where $M,N\dots$\ run from $1\dots D$,    $a,b\dots$ run over $  0\ 
\mbox{and } 1$,  with $\epsilon^{01}=1$ and 
 $\xi ^a$ are the string worldsheet coordinates. The spacetime
coordinates are denoted by $X^{M}$ and 
$\gamma$ is the induced metric on $\Sigma$, the string worldsheet. The
background spacetime metric and antisymmetric tensor fields
in which the string propagates are denoted by $g$ and $B$
respectively.

In bosonic string theory there is an additional  massless mode to
consider, namely  the dilaton, $\Phi$.
This gives a contribution to the action of the form

\ben \label{dil}
S_{dil} = -{1\over{8\pi}}\int d^2\xi{\sqrt \gamma}R^{(2)}\Phi (X) ,
\een
where $R^{(2)}$ is the scalar curvature of the metric $\gamma$.
This term breaks Weyl invariance on a classical level as do the one 
loop corrections to $g$ and $B$. Given a set of background fields we find that conformal invariance is restored to first order in $\alpha'$ provided that the string 
$\beta$-function equations are satisfied \cite{cfmp}

\ben \label{betafns} 
R_{MN}-\nabla_M\nabla_N\Phi -{1\over 4}H_M^{\ LP}H_{NLP}&=&0 \\
\nabla_LH^L_{\ MN}+ (\nabla_L\Phi)H^L_{\ MN}&=&0 \\
R-{{D-26+c}\over{3\alpha'}}-(\nabla\Phi)^2 - 2\Box\Phi - {1\over 12}H_{MNP}H^{MNP}&=&0
\, . 
\een
for $H=3dB$, $c$ is a central charge and $D$ is the
dimension of the target space.
  
If the fields represent an exact conformal field theory there are no higher order
corrections to the beta functions. This would be the case for exact
string vacua, such as those for plane waves. It is these conformal field theories that we wish to consider. After the
manipulations have been performed exactly, it is then possible to find
the true low
energy limit of the dual theory.

\section{Gauged $WZNW$ models}
\news

\subsection{$WZNW$ models}  
One of the simplest ways to construct a conformal field theory
is to consider a $WZNW$ model for a group $G$ \cite{wit,jjmo,ag}. If the
group $G$ has a
subgroup $H$ then a new model can be constructed by gauging the $H$ 
symmetry of the $\sigma$-model. This is the basic idea of T-duality.

 We consider the $WZNW$ action for a group $G$ with subgroup $H$ and
elements $g(z,\bar z)\in G$

\ben\label{WZNW}
I(g)&=&{1 \over 4\pi} \int_{\partial B} d^2z\  Tr\left(g^{-1}\partial g .g^{-1} {\bar
{\partial}} g \right)-{1\over{12\pi}}\int_{B}dV\
Tr\left(g^{-1}dg\wedge g^{-1}dg\wedge g^{-1}dg \right)
\nonumber \\
S(g)&=&kI(g)\,,
\een 
where $d$ is the exterior derivative operator, k is an integer called the {\it{level}} of the model, $V$ is the volume element
for $B$, and the boundary of $B$, $\partial B$, is the flat 2D string
worldsheet $\Sigma$. The second term in the action is called the Wess-Zumino-Novikov-Witten
term and is an integral over any three-manifold with boundary
$\partial B$. The group element $g$ is extended to be a map from $B$
to $\diff B$. We employ  complex worldsheet
coordinates $z,\bar z$ 
\be
z=\frac{\xi_1+i\xi_2}{\sqrt {2}},\ \ \bar{z}=\frac{\xi_1-i\xi_2}{\sqrt
{2}}\,
\ee

 The model is manifestly invariant under a global $G_R\times
G_L$ group of transformations, and the coefficients of the WZNW term have been chosen such that
the equations of motion 
\be\label{eom}
\partial\left({\bar{\partial}g}.g^{-1}\right)={\bar{\partial}}\left(g^{-1}\partial
g\right)=0\,
\ee
hold \cite{wit}. This only occurs at a fixed point in the
renormalisation group flow for the case of arbitrary coupling of the
WZNW term to the action (\ref{WZNW}), and implies that the group
elements are separable into holomorphic and anti-holomorphic parts
\be
g=h(z)f(\bar{z})\,.
\ee 
 The model then has the property, called the `Polyakov-Wiegmann Theorem' \cite{pw}, that
\be\label{WZNWid}
I(gh)=I(g)+I(h)-{1\over{2\pi}}\int_{\partial B} d^2z\  Tr\left(g^{-1}\partial g  .{\bar
{\partial}} h h^{-1} \right)\,.
\ee 
If $G$ has a subgroup $H$ then the $H$ symmetry can be gauged using
this property to give
a new model \cite{witbh}. Before
performing the gauging procedure, we need to investigate carefully
the quantum mechanical properties of the action. In order to define
the quantum effective action we perform a Legendre
transformation \cite{ts}. The result is that the classical action is
equivalent to the quantum action if we make the transformation of the
level $k$ as follows
\be\label{leveltrans}
k\rightarrow k+{1\over 2}c_G\,,
\ee
where $c_G$ is the dual Coxeter number of the defining group
$G$. This
is given by
\be
c_G={{f_{a}}^{bc}}{f^{a}}_{bc}\,,
\ee
where the ${{f_{a}}^{bc}}$ are the structure constants of the Lie
algebra of the group $G$, the indices of which  are
raised and lowered using the Cartan metric $\eta^{ab}$ on $G$\footnote{Clearly, in the abelian case, this quantum mechanical
shift vanishes and the transformation is trivial.}.
This result may also be obtained from a current algebra point of view
where it can be shown that the Sugawara energy momentum tensor has
the
renormalisation factor \cite{kz} arising from the normal ordering
ordering of the currents, denoted by  $\stackrel{\mbox{\tiny o
\normalsize}}{\mbox{\tiny o \normalsize}}$ $\stackrel{\mbox{\tiny o \normalsize}}{\mbox{\tiny o \normalsize}}$ 
\be
T(z)=\frac{1}{k+{1\over 2}c_G} \stackrel{\mbox{\tiny o
\normalsize}}{\mbox{\tiny o \normalsize}} J(z)^2 \stackrel{\mbox{\tiny
o \normalsize}}{\mbox{\tiny o \normalsize}} \,
\ee

 Thus in order to obtain the consistent quantum mechanical
action we must alter the level of the $WZNW$ model by
${1\over2}c_G$. This means that the target space metric is given by
the group space metric multiplied by $k+{1\over 2}c_G$. Although this
is a trivial observation for the case of a standard $WZNW$ model defined by a
group $G$, the shift enters in a non-trivial way for a gauged model,
as we shall see in the next section.

\bigskip
\subsection{Vector gauging of the $\sigma$-model}

 We now gauge the $H$ symmetry from the action
(\ref{WZNW}). Under a general $H$ gauging we have
\ben
g&\rightarrow& u_Lgv_R \nonumber \\
u_L,v_R&\in& H,
\een

 If $H$ is abelian then both the axial-vector and the vector gaugings are
anomaly free \cite{kiritsis}. For the case where $H$ is a non-abelian group only the
vector gauge is believed to be anomaly free. We thus wish to 
consider actions which are invariant under the vector gauging, which
corresponds to the case $u_L=v_R^{-1}$
\be
g\rightarrow u(z,\bar{z})gu^{-1}(z,{\bar{z}})\,.
\ee

 Substituting these transformations into (\ref{WZNW}) and using
the identity (\ref{WZNWid}) we find that 
\ben\label{sa}
I(g,A,{\bar A})&=&I(g)+\int d^2z\
Tr\left(A\bar{\partial} g g^{-1}-\bar{A}g^{-1}\partial
g+A\bar{A}-g^{-1}Ag\bar{A}\right)\nonumber \\
&=&I(g)+I_1(g,A,\bar{A})\,,
\een
with appropriate gauge fixing for $g$.
In order to leave the dynamics unchanged we may add a Lagrange
multiplier term to each side of this equation:

\be\label{lag}
I_{L}=\frac{1}{4\pi}\int d^2z{Tr\left(\chi F\right)}\,,
\ee
where $\chi$ is the Lagrange multiplier term and the field strength
of the gauge field is given by

\be
F=\diff{\bar{A}} -\bdiff A +[A,\bar{A}]\,.
\ee
The equations of motion for $\chi$ imply a vanishing $F$, and
we may thus parametrise the gauge fields by 
\be
A=h^{-1}\partial h\hskip1cm \bar{A}=\bar{h}^{-1}\bar{\partial}\bar{h}\hskip1cm
u=h{\bar h}\,.
\ee

 The action $S(g,A,\bar{A})=kI(g,A,\bar{A})$ is then invariant under the vector
gauge transformations 

\ben\label{vectorgauge}
g&\rightarrow& ugu^{-1}\nonumber \\
A&\rightarrow& u\left(A+\partial\right)u^{-1}\nonumber \\
\bar{A}&\rightarrow& u\left(\bar{A}+\bar{\partial}\right)u^{-1}\,.
\een

\subsection{Quantum effective action}

We now consider the quantum mechanics of the gauged WZNW
model,neglecting for the moment the Lagrange multiplier term, with
bare $k$. This is defined by the  path integral

\be\label{z}
Z=\int[dg][dA][d\bar{A}]\exp\left(-kI(g)-kI_1(g,A,\bar{A})\right)\,,
\ee
 The measures given are the gauge fixed group invariant canonical
measures.
 
\medskip

 From the previous discussion we must shift the level
$k$ in order to have a quantum mechanically consistent action. To attempt to do
this we must try to rewrite the gauged action (\ref{sa}) in terms of $WZNW$
models.

 We note that the identity (\ref{WZNWid}) gives us
\be\label{id}
I(g,A)=I(h^{-1}g\bar{h})-I(h^{-1}\bar{h})\,
\ee
thus, changing variables in the path integral using the identity
\be
\det\left(\partial+[A,\ ]\right)\det\left(\partial+[\bar{A},\ ]\right)=\exp\left(c_HI(h^{-1}\bar{h})\right)\det\partial\det\bar{\partial}\,
\ee
and substituting (\ref{id}) we obtain the new, equivalent, path integral 
\be\label{newz}
Z=\int[dg][dh][d\bar{h}]\exp\left[-kI(h^{-1}g\bar{h})+(k+c_H)I(h^{-1}\bar{h})\right]\,.
\ee

 This implies that the system can be quantised as the sum of
two separate $WZNW$ models. Since $h^{-1}G\bar{h}=G$ and $h^{-1}H=H$, we see that the first term in (\ref{newz}) is
a $WZNW$ model for group $G$ with level $k$ and the second term is a $WZNW$
model for the group $H$ with level $-(k+c_H)$.

 We now use the result that we must shift $k$ by half the
dual Coxeter number of the group in question. This gives us the effective
quantum action 
\be
\Gamma(g,A)_{h,\bar{h}}=(k+{1\over 2}c_G)I(h^{-1}g\bar{h})-(k+{1\over
2}c_H)I(h^{-1}\bar{h})\,.
\ee

\medskip
 Performing the reverse change of variables and rewriting the
action in terms of $A,\bar{A}$ we arrive at the true quantum
mechanically consistent path integral

\ben\label{effact}
Z&=&\int[dg][dA][d\bar{A}]\exp\left(-\Gamma(g,A,\bar{A})\right)\nonumber
\\
\Gamma(g,A,\bar{A})&=&(k+{1\over2}c_G)\left(I(g,A)+\frac{c_G+c_H}{2(k+{1\over 2}c_G)}I(h^{-1}\bar{h})\right)\,.
\een

 For large $k$ the first part of this action is the dominant
term and represents the low energy part of the action. The second term
becomes important at small $k$ and is the quantum correction to the
classical action.

\bigskip
\subsection{The dual model}

\medskip
 The dual model is obtained by integrating out the
$A,\bar{A}$ dependence from the path integral (\ref{effact}). This is
done using Gaussian path integral techniques. We write the components
of the gauge field as $A=A_aT^a$ for $T^a$ the generators of the
subgroup $H$ in the adjoint representation. Rearranging the terms in the action and using
the identity (\ref{id}) we obtain 
\ben\label{newact}
\Gamma(g,A,\bar{A})&=&(k+{1\over2}c_G)\left(I(g)+I_L+{1\over{4\pi}}\int
d^2z \ Tr\left(M_1+M_2+M_3\right)\right)\nonumber \\
M_1&=&A_aT^a\bar{\partial} g g^{-1}-\bar{A}_bT^bg^{-1}\partial
g- A_a\bar{A}_bg^{-1}T^agT^b\nonumber \\
M_2&=&\left(1-\frac{c_G+c_H}{2(k+{1\over2}c_G)}\right)A_a\bar{A}_bT^aT^b\nonumber
\\ 
M_3&=& \left(\frac{c_G+c_H}{2(k+{1\over
2}c_G)}\left(I(h^{-1})+I(\bar{h})\right)\right)\,.
\een

We consider the dual description of the Lagrange multiplier term $I_L$
obtained by integrating by parts.  This removes the derivatives of $A$ and introduces derivatives
of $\chi$ into the action
\be{\label{lm}}
I_{L}=\frac{1}{2\pi}\int
d^2z Tr\left(-\bar{A}^a\diff\chi_a+A^a\bdiff\chi_a+A_c\bar{A}_d{f_a}^{cd}\chi^a\right)\,,
\ee
where $A=A_aT^a, \chi=\chi_aT^a$ for $T^a$ the generators of the
subgroup $H$ in the adjoint representation.

 The terms $I_L$, $M_1$ and $M_2$ are now each  local quadratic in
$A,\bar{A}$, and the  gauge field dependence can be integrated out  using results for Gaussian
integrals, giving the contribution to the path integral

\ben\label{zzz}
Z&=&\int[dg]\exp(-\hat{S})\nonumber
\\
&=&\int[dg]\exp\left(-(k+{1\over 2}c_G)\hat{I}(g,T^a)+J\right)\nonumber
\\
\hat{I}(g)&=&{1\over 4\pi}\int d^2z \ \left(\bdiff\chi^a+Tr \left( T^a{\bar
\partial}gg^{-1} \right)\right) \left( \Lambda\right)_{ab}^{-1}
\left(-\diff\chi^b+Tr
\left( T^b g^{-1} \partial g \right)\right)\nonumber \\
\Lambda_{ab}&=&\chi_a {f^a}_{bc}+Tr\left(\omega(k)T_a T_b - T_a gT_b g^{-1}
\right)\nonumber \\
\omega(k)&=&1-\frac{c_G+c_H}{2(k+{1\over2}c_G)}\,,
\een

 $J$ is an anomalous Fadeev-Popov determinant factor obtained when
integrating out the gauge field, which can be evaluated using standard
regularisation techniques and is given by \cite{buscher} 

\be\label{j}
J=\frac{1}{8\pi}\int d^2z\ R^{(2)}\log(\det \Lambda))\,.
\ee
From the path integral \ref{zzz}  we can read off the contributions from
$I_L,M_1$ and $M_2$ to the new dual action $\hat{S}$.

This is the form of the action obtained under the usual non-abelian
duality procedure \cite{t} if we take the large $k$ limit. It is not
necessarily 
an exact expression because we must consider the effect of the
anomalous term $M_3$, which may not vanish, even in the low energy
limit.
 Although the term is cancelled out in the
classical 
form of the action, it survives in the quantum action due to the
shift in the Coxeter numbers for the case of a non-abelian group. We discuss the effects of this term in
the next section.

\subsection{Anomaly term}

We must now discuss the term $M_3$, which is the  obstacle
to integrating out the gauge field $A$ completely. We will show that
if the gauged subgroup is semisimple, then we may integrate out  the gauge
field dependence  from  $M_3$, enabling us to give the dual solution for
any $k$. 

To begin with we define \cite{ts}

\be
F':=\bar{\partial} A+\partial{\bar A}
\ee

and consider
\be
F'{1\over{{\bar\partial}\partial}}F' =
\bdiff A{1\over \diff}A +\diff {\bar A}{1\over \bdiff}{\bar A} +\diff
{\bar A}{1\over \diff}A + \bdiff A {1\over \bdiff}{\bar A} \,,
\ee
where the inverse ${1\over \diff}$ is the Greens function for the $\diff$
operator. Using the cyclic property of $Tr$, and implicit
integrations throughout we find that 
\be
\int d^2z Tr \left(F'{1\over{{\bar\partial}\partial}}F'+2A\bar{A}\right)=-\int d^2z Tr \left(A{\bdiff\over \diff}A+{\bar A}{\diff\over \bdiff}{\bar A}\right)\,
\ee

\medskip
We now relate this to the anomalous term $M_3$, which is of the form

\ben M_3\label{M3} &\propto&
I(h^{-1})+I({\bar h}) \nonumber \\ &=&{1\over 4\pi}\int_{\diff B}d^2z Tr \left(
h^{-1}\diff h.h^{-1}\bdiff h+ {\bar h}^{-1}\diff {\bar h}.{\bar
h}^{-1}\bdiff {\bar h} \right)\\ \nonumber &\  &\ \ \ \ \ \ + \mbox{WZ term}\,.
 \een

We attempt to write this integral in terms of the field $F'$. First we see that

\ben A{\bdiff\over \diff}A&=&\left(h^{-1}\diff h\right){\bdiff\over
\diff}\left(h^{-1}\diff h\right)\nonumber\\
&=&\left(h^{-1}\diff h\right){1\over \diff}\left(\bdiff h^{-1}.\diff h
+ h^{-1}\bdiff \diff h \right)\nonumber \\
&=&\left(h^{-1}\diff
h\right){1\over\diff}\left(\diff\left(h^{-1}\bdiff
h\right)+\left[\bdiff h^{-1}.h,h^{-1}\diff h\right]\right) \nonumber \\
&=&h^{-1}\diff h\left(h^{-1}\bdiff h+J\right)\,
\een
where 
\be 
J={1\over \diff}\left(\left[h^{-1}\diff h,h^{-1}\bdiff
h\right]\right)\,.
\ee

We note that there are no surface terms because $\diff B$, being the
boundary of $B$, has no boundary, hence

\be
 h^{-1}\diff h.h^{-1}\bdiff h = A{\bdiff\over\diff}A-h^{-1}\diff h.J\
\ee

Repeating the procedure for the boundary term in $\ref{M3}$ we obtain
\be
-A{\bdiff\over \diff}A-{\bar A}{\diff\over \bdiff}{\bar A}=
 \diff h^{-1}.\bdiff h + \bdiff {\bar h}^{-1}\diff{\bar h}
-K -\bar{K}\,,
\ee
where we have defined
\ben
K&=&\left(h^{-1}\diff h \right) J\nonumber \\
\bar{K}&=&\left({\bar h}^{-1}\bdiff {\bar h}\right){\bar J}\,.
\een

Clearly in the abelian case the non-local terms $J,{\bar J}$
vanish. For the non-abelian case the commutator will be in general
non-zero. We  look for      
            special cases where the integrals may be evaluated.

Substituting in the expression for $M_3,\  \ref{M3},$ we obtain 

\be\label{faction}
I(h^{-1})+I({\bar h})={1\over 4\pi}\int_{\diff B}d^2z Tr
\left(-2A{\bar A}-F'{1\over \bdiff\diff}F'-(K+{\bar K})\right)+(WZ)\,
\ee

We now consider the parameterisation of the group elements.
The generators of the Lie algebra obey
\ben
A&=&A_{a}T^a:\left[ T^a,T^b \right]={f_{c}}^{ab}T^c\\
Tr(T^aT^b)&=&2\eta^{ab}\,,
\een
where $\eta^{ab}$ is the Cartan matrix for the group, which is used to
raise group indices. This implies that the group elements are locally given by
\be
h=\exp{A_a(z,{\bar z})T^a}\hskip1cm
{\bar h}=\exp{{\bar A}_a(z,{\bar z}){\bar T}^a}\,
\ee
The bar denotes complex conjugate and we treat $z$ and ${\bar z}$ as
independent variables on the two-sphere.
This implies that
\be
\left[h^{-1}\diff h,h^{-1}\bdiff h\right]=\diff
A_a\bdiff A_b{f_{c}}^{ab}T^c\,
\ee

Expressions of this form will hold in any local neighbourhood on the
group. This then leads to the term $J$ being of the form
\ben
J&=&{1\over \diff}\left( {f_{c}}^{ab}T^c \diff A_a\bdiff A_b \right) \\
&=&{f_{c}}^{ab}T^c\int dz' G_{z}(z,z')\diff A_a(z',{\bar z})\bdiff A_b(z',{\bar z}) \\
&=&{f_{c}}^{ab}T^c \Lambda_{ab}\,,
\een
for some functions $\Lambda_{ab}$.

The contribution to $I(h)$ from $J$ is now of the from
\be
I_J   \alpha\int_{\diff B}d^2z Tr \left( \diff A_dT^d
{f_c}^{ab}T^c\Lambda_{ab}\right)\,
\ee

Applying the normalisation condition $Tr(T^aT^b)=2\eta^{ab}$ for the
generators of the Lie algebra we obtain
\ben\label{igeneral}
I_J&\propto&\int_{\diff B}d^2z  \left( \diff A_d\eta^{cd}
{f_c}^{ab}\Lambda_{ab}\right)\\
&=&\int_{\diff B}d^2z  \left( 
f^{dab}\diff A_d\Lambda_{ab}\right)\,.
\een
Recall that we define $f^{dab}=\eta^{dc}{f_c}^{ab}$.

In general this expression cannot be simplified, so we restrict
ourselves to the special case where the coefficient of the WZ
functional is chosen as in
$\ref{WZNW}$ to
make the equations of motion hold. These equations of motion imply
that the group elements are separable in terms of the worldsheet
coordinates,

\ben
h(z,{\bar z})&=&p(z)q({\bar z})\\
\Rightarrow h&=&\exp (P_a(z)T^a)\exp (Q_b({\bar z})T^b)\\
\Rightarrow h^{-1}\diff h&=&q^{-1}\diff P_aT^aq\\
h^{-1}\bdiff h&=& {\bar \diff}Q_aT^a\,
\een

We then obtain the contribution to $I$ of the form
\be
I_J  \propto\int_{\diff B}d^2z Tr\left( q^{-1}P_dT^dq{1\over \diff}\left( \left[ q^{-1}P_aT^aq,Q_aT^a\right] \right)\right) \,
\ee

Of course, the group elements $p$ and $q$ commute with the products
$P_aT^a$ and $Q_aT^a$ respectively, being polynomials in these
terms. Making use of this fact, and using the trace in the integral,
we obtain
\ben
I_J&\propto&\int_{\diff B}d^2z Tr\left(\diff P_dT^d{1\over \diff}\left( \left[
\diff P_aT^a,\bdiff Q_bT^b\right] \right)\right) \\
&=&\int_{\diff B}d^2z Tr\left(\diff P_dT^d\bdiff Q_b{f_c}^{ab}T^c {1\over
\diff}\left(\diff P_a\right)\right)\,
\een
We now take the trace over the two remaining matrix elements to obtain
\ben
I_J&\propto&\int_{\diff B}d^2z \left(\diff P_d\bdiff Q_bf^{dab} {1\over \diff}\left(\diff P_a\right)\right)\\
&=&2\int_{\diff B}d^2z \left( \diff P_d\bdiff Q_b P_af^{dab}\right)\\
&=&0\,, 
\een
by integration by parts.

We now need to consider the Wess Zumino terms in $\ref{M3}$.
These are 

\ben 
WZ&\propto&\int_{B}
d^3z Tr\left( h^{-1}dh\wedge h^{-1}dh\wedge h^{-1}dh-{\bar
h}^{-1}d{\bar h}\wedge {\bar h}^{-1}d{\bar h}\wedge {\bar h}^{-1}d
{\bar h}\right)\nonumber \\
&=&{1\over 2}\int_{\diff B} d^2z \epsilon^{\mu\nu}\diff_\mu A_a\diff_\nu A_b
b^{ab}-(A\leftrightarrow{\bar A})\nonumber \\
&=&0\nonumber \\
&\iff& T^aT^bT^c\equiv H^{abc}=\frac{1}{2}(db)^{abc}\\
&\iff& f^{abc}=f^{[abc]}\nonumber \,
\een

We now make use of the fact that in the adjoint
representation, the structure constants are completely anti-symmetric 
for all compact, $\it{semi simple}$ Lie algebras in an
orthonormal basis$\cite{groups}$. We thus have the vanishing of the extra anomalous WZ term in the
cases where the gauged subgroup $H$ is one of $SO(N)$, $SU(N)$, $Sp(N)$,
$E_6$, $E_7$, $E_8$, $F_4$, $G_2$.

\bigskip

Recall that we are trying to simplify the term $\ref{faction}$. The
final part to consider is the $F'$ term. To complete the analysis we note that we are free to pick a gauge for
the problem. 
\ben\label{fgauge}
F'&=& \diff{\bar A}+{\bdiff A}\\
F'_{gauged}&=&\diff(A+\bdiff \Lambda)+\bdiff (A+\diff \Lambda)\\
&=& 0\nonumber \\
&\iff& 2\diff \bdiff \Lambda = \diff {\bar A}+ \bdiff A\,.
\een
The equation $\diff \bdiff \Lambda = S$ has a solution if 
\be
\int_{\diff B} S=0\,,
\ee
This is identically true for $S=\diff {\bar A}+ \bdiff A$, therefore we
may always choose a gauge to set the $F'$ term to zero.

We now have the result that for a semi-simple (or abelian) Lie group, the
anomalous term $M_3$ can be reduced to a term involving only $A, {\bar A}$,
which can be simply integrated out of the path integral. 

\be
I(h^{-1})+I({\bar h})={1\over 4\pi}\int_{\diff B}d^2z Tr
\left(-2A{\bar A}\right)\,
\ee

This term simply alters the premultiplying factor in $M_2$, doubling
the $k$-dependent term. The final action is given by
\ben{\label{finalaction}}
S_{dual}(g)&=&I(g)+\left( \frac{k+\frac{1}{2}c_G}{4\pi}\int d^2z \left(\bdiff\chi^a+Tr \left( T^a{\bar
\partial}gg^{-1} \right)\right) \left( \Lambda\right)_{ab}^{-1}
\left(-\diff\chi^b+Tr
\left( T^b g^{-1} \partial g \right)\right)\right)-J\nonumber \\
\Lambda_{ab}&=&\chi_a {f^a}_{bc}+Tr\left(\omega(k)T_a T_b - T_a gT_b g^{-1}
\right)\nonumber \\
J&=&\frac{1}{8\pi}\int d^2z\ R^{(2)}\log(\det \Lambda))\\
\omega(k)&=&1-\frac{c_G+c_H}{(k+{1\over2}c_G)}\,,
\een

 The resulting action, which is valid for the gauging of any
semi-simple or abelian subgroup, can be interpreted as
the effective quantum action and is exact for all $k$. The gauged
action is therefore conformally invariant to all loop orders.

\subsection{Quantum Corrections to the gauged theory}

Taking into account the quantum properties of the dual action we
find that there are two different anomalies which occur under the
gauging procedure. 

\begin{itemize}
\item Under the gauging procedure we saw that there was a gauge fixing anomaly term which arises
in the Fadeev-Popov determinant in the path integral. This term is proportional to the scalar
curvature of the worldsheet of the form (\ref{dil}) and spoils the conformal invariance of
the $\sigma$-model at any energy scale. We may not, therefore, neglect
this term. We see, however, that the anomaly contribution is of exactly the same form as the
dilaton coupling term in the string theory (\ref{dil}) and we can therefore
counteract the effect of the gauge fixing by introducing a dilaton
term into the action, or we can consider the transformation to induce
a dilaton shift. This causes the anomaly to vanish, and the original
and dual string theories are equivalent.

\item The level $k$ enters into the expressions for the
dual model \ref{finalaction} in a non-trivial way. These $k$ dependences are higher order quantum
corrections to the dual action. In the case where the gauged group is
semisimple as $k$ becomes very large,
corresponding to a low energy string in the given background, we may
neglect the higher order terms in $1\over k$. The quantum corrections vanish
and we are left with the classical expression for the gauged
theory. We may
therefore neglect the $k$ dependence at low energies in these circumstances.

\end{itemize}

\subsection{Application to string theory}

 We can interpret the previous results in terms of string theories by treating
the boundary $\partial B$ of the Wess-Zumino term as the string worldsheet. Since the
$WZNW$ model is a bosonic, massless theory for the choice of constants we have used, it is
consistent to consider only the bosonic, massless degrees of freedom
of the string: the metric, the antisymmetric tensor field and
the dilaton. We can then
interpret the group space metric of the WZNW model to be the spacetime
metric $g$ and the $WZ$ term to be the antisymmetric tensor field $B$.
 There is no dilaton term for the
ungauged WZNW model. The level $k$ can be associated with the free
parameter of string theory. Since string
theorists work with the inverse string tension we relabel $k\rightarrow {1\over
\alpha'}$ for convenience obtaining

\ben \label{a1} 
S&=&{1\over{4\pi\alpha'}} \int d^2z Q_{NM}\partial X^N {\bar \partial}X^M\nonumber \\
Q&:=&g+B\,.
\een
Under the duality procedure (gauging a subgroup of the symmetry and
then integrating out the gauge fields
$A,\bar{A}$) we obtain the action for the dual WZNW model,
$S_{dual},\ref{finalaction}$ from which we can read off
the dual string modes, provided that the gauged subgroup is
semisimple. This differs from the standard non-abelian duality formula
by the $k$ dependance in the $\omega(k)$ term

\be
\omega(k)=1-\frac{c_G+c_H}{(k+{1\over2}c_G)}\,.
\ee

As we take large $k$ we note that $\omega(k)\rightarrow 1$, and the
expression \ref{finalaction} reduces to the usual non-abelian duality
formulae. Thus,  
in the low energy limit the result is the same as that
obtained via the usual non-abelian target space duality procedure. The
work explains the  result $\cite{t}$ that for low
energy string theory  conformal invariance may be restored in the dual model
by a shift in the dilaton in the case that the gauged subgroup is
semisimple. In the non-semisimple case the
effect from the non-local anomaly term does not necessarily
vanish for large $k$ and thus the expressions for the new fields may
depend non-trivially on $k$, even at low energy. In these situations
the dual  model does not have a background spacetime interpretation. Thus
there is no dual string background which satisfies the string $\beta$
function equations. This behaviour had been observed when the string
T-duality equations were applied to models with non-semisimple
structure constants \cite{t}.

\section{Conclusion}

We have discussed non-abelian gaugings of WZNW models. We are able to present
exact dual models for arbitrary values of the level $k$ in the case
where the gauged group is semisimple.

 If we interpret the models in terms of string theories
then in the large $k$ limit, corresponding to small string
tension, we reproduce the standard non-abelian duality formulae with
an additional $k$-dependant  anomaly term. This term can be shown to vanish in the
low energy limit in the semisimple case. In these situations  this
shows that the standard string  non-abelian target space duality 
 is well defined quantum mechanically. 

If the gauged subgroup is not semisimple then the anomaly term does
not necessarily become negligible  at large $k$, and is not of the
form of the dilaton factor.  In this
situation, although we may construct the  dual model, the resulting
CFT does not 
have a low energy background spacetime  interpretation. Thus the
string duality formalism breaks down. 

To conclude, we have shown that in the semisimple case non-abelian target space duality is a
perturbatively exact quantum symmetry of the string theory. As such,
its role in string theory must be given serious consideration.

\pagebreak
\vfill\eject

\end{document}